\documentclass[twocolumn]{aastex63}

\usepackage{amsmath}
\usepackage{amssymb}
\usepackage{threeparttable}
\usepackage{makecell}
\usepackage{longtable}
\usepackage{latexsym}
\usepackage{multirow}
\usepackage{CJKutf8}
\usepackage{color}

\received{XXX}
\revised{YYY}
\accepted{ZZZ}
\submitjournal{ApJ}
\shortauthors{Liu et al.}

\begin{document}
	
	\begin{CJK*}{UTF8}{gbsn}
		\end{CJK*}

	\title{Coherent Cherenkov Radiation by Bunches in Fast Radio Bursts}
		
		\author[0000-0002-5758-1374]{Ze-Nan Liu}
		\affiliation{School of Astronomy and Space Science, Nanjing University, Nanjing 210023, China}
		\affiliation{Key Laboratory of Modern Astronomy and Astrophysics (Nanjing University), Ministry of Education, Nanjing 210023, China}
        
				\author[0000-0001-9648-7295]{Jin-Jun Geng}
		\affiliation{Purple Mountain Observatory, Chinese Academy of Sciences, Nanjing 210023, China}
  
		       	\author[0000-0001-6374-8313]{Yuan-Pei Yang}
		\affiliation{South-Western Institute for Astronomy Research, Yunnan University, Kunming 650500, China}
		\affiliation{Purple Mountain Observatory, Chinese Academy of Sciences, Nanjing 210023, China}
  
		\author[0000-0001-9036-8543]{Wei-Yang Wang}
		\affiliation{School of Astronomy, University of Chinese Academy of Sciences, Beijing 100049, China}
		\affiliation{Department of Astronomy, Peking University, Beijing 100871, China}
		
		\author[0000-0002-7835-8585]{Zi-Gao Dai}
		\affiliation{Department of Astronomy, University of Science and Technology of China, Hefei 230026, China; daizg@ustc.edu.cn}

		\begin{abstract}		
			Fast radio bursts (FRBs) are extragalactic radio transients with extremely high brightness temperature, which strongly suggests the presence of coherent emission mechanisms. In this study, we introduce a novel radiation mechanism for FRBs involving coherent Cherenkov radiation (ChR) emitted by bunched particles that may originate within the magnetosphere of a magnetar. We assume that some relativistic particles are emitted from the polar cap of a magnetar and move along magnetic field lines through a charge-separated magnetic plasma, emitting coherent ChR along their trajectory. The crucial condition for ChR to occur is that the refractive index of the plasma medium, denoted as $n_r$, must satisfy the condition $n_r^2 > 1$. We conduct comprehensive calculations to determine various characteristics of ChR, including its characteristic frequency, emission power, required parallel electric field, and coherence factor.  Notably, our proposed bunched coherent ChR mechanism has the remarkable advantage of generating a narrower-band spectrum. Furthermore, a frequency downward drifting pattern, and $\sim100\%$  linearly polarized emission can be predicted within the framework of this emission mechanism.                 
		\end{abstract}
		
		
		\section{Introduction}
Fast radio bursts (FRBs) are transient radio signals lasting mere milliseconds  \citep{Lorimer2007,Keane2012,Thornton2013}, their physical origin and radiation mechanism remain unknown. To date, hundreds of FRBs have been detected, with only a small fraction displaying repeating patterns\footnote{A Transient Name Server system for newly reported FRBs \citep[https://www.wis-tns.org,][]{Petroff2020}.}. Recently, \cite{Bochenek2020} and \cite{CHIME2020b} reported a bright FRB-like burst from a Galactic magnetar SGR J1935+2154, suggesting that magnetars are capable of producing at least some FRBs. Many theoretical models have been proposed to explain FRBs \citep[see details from ][]{Platts2019,ZB2020Natur,XWD2020,Lyubarsky2021}. Based on their spatial location, these bursts are primarily categorized into two categories: close-in models, which occur within the neutron star's magnetosphere \citep[e.g.,][]{Kumar2017,Yang2018,Yang2021,Lu2020,Wang2019,Wang2020}, and far-away models, which are situated outside the neutron star's magnetosphere \citep[e.g.,][]{Lyubarsky2014,Beloborodov2019,Metzger2019,Margalit2020}. Besides, \cite{Lyubarsky2020} proposed an intermediate model in which the emission is generated through the compression of the striped wind near the light cylinder by a fast magnetosonic (FMS) wave. This model was subsequently tested by \cite{Mahlmann2022} using particle-in-cell simulations.

The high brightness temperature of FRBs implies the necessity of coherent radiation  \citep{Lorimer2007, ZB2020Natur}. Coherent curvature radiation (CR) occurs when relativistic particles travel along curved magnetic field lines. Some models invoking coherent curvature emission by charged bunches can explain the radio emission of pulsars  \citep{Ruderman1975,Sturrock1975,Elsaesser1976,Cheng1977,Melikidze2000,Gil2004,Gangadhara2021} and FRBs \citep{Katz2014,Kumar2017,Yang2018,Ghisellini2018,LuKumar2018,Katz2018,WangLai2020,Wang2020,Cooper2021,WangYang2021,liu2023}. Additionally, \cite{ZB2022} proposed a model invoking coherent inverse Compton scattering (ICS) by bunches as the radiation mechanism of FRBs. Similar to the coherent CR, the coherent ICS mechanism is not intrinsically dependent on the dispersive properties of the plasma.

The relativistic plasma emission (plasma maser) is also a possible radiation mechanism. Langmuir-like waves are generated by a beam instability between a beam of relativistic particles and a background plasma, and then part of the wave energy is converted to escaping radiation \citep{Melrose1991}. The growth rate of beam instability depends on the dispersive properties of the plasma. However, the growth rates for beam instabilities are too small to be effective \citep{Melrose2017}. The interaction between the beam and the background plasma can potentially generate solitons through alternative mechanisms. ChR is generated when a relativistic charged particle sweeps through a medium, but this occurs only if the particle's velocity surpasses the phase velocity of light in that medium \citep{Jelley1958,jackson1998}. ChR arises from the coherence of the emitted secondary wave. The electrons generating the secondary wave within the medium experience acceleration as the field of the ``superluminal'' particle sweeps through the medium. The coherent ChR mechanism has been explored to interpret many astrophysical phenomenas. The coherent radio emission from high-energy showers in the dense media was proposed as a possible way to detect high-energy cosmic rays \citep{Askaryan1962}. Furthermore, \cite{Alvarez2012} successfully simulated the radio emission from the extensive air showers by the Monte Carlo approach. Besides, the ChR from relativistic charged particles moving through a charge-separated magnetic plasma was discussed by \cite{Kolbenstvedt1977} to explain the observed radio emission from the Crab pulsar. However, detailed discussions regarding the spectrum, total radio luminosity, polarization, and other aspects have not been provided.

In this paper, we propose a radiation mechanism of FRBs invoking coherent ChR emitted by bunched particles that may be generated within the magnetosphere of a magnetar. Assuming that some relativistic particles are emitted from the polar cap of a magnetar and move along magnetic field lines through a charge-separated magnetic plasma, emitting coherent ChR along their trajectory. The paper is organized as follows. We first discuss the coherent ChR produced by accelerated charges within the magnetar magnetosphere in Section \ref{sec2}, including the magnetized plasma properties (Section \ref{sec2.1}), ChR power  (Section \ref{sec2.2}),  required electric field (Section \ref{sec2.3}), and coherence factor (Section \ref{sec2.4}). The observation tests of the emission mechanisms are shown in Section \ref{sec3}, including the spectrum (Section \ref{sec3.1}), frequency drifting (Section \ref{sec3.2}), and polarization (Section \ref{sec3.3}). The results are discussed and summarized in Section \ref{sec4}. The convention $Q_x = Q/10^x$ in cgs units is used throughout this paper.

\section{Coherent ChR from Accelerated Charges}
\label{sec2}

 The electrons generate the secondary wave within the medium experience acceleration when the field of the  ``superluminal'' particles sweeps through the medium. Therefore, the ChR is essentially the collective effect of numerous particles interacting within the medium. We assume that an ultrarelativistic primary beam is emitted from the polar cap and moves along the field lines through a relativistic electron-positron plasma produced by the pair cascade process.This process leads to the ChR along the trajectory. The Cherenkov angle $\theta_c$ between radiation direction and particle motion direction can be given by 
\begin{equation}
\cos\theta_c=\frac{u}{v}=\frac{c}{n_r v},
\end{equation}
where $u=c/n_r$ is the speed of light in the medium, $n_r$ is the refractive index. The faster the particle speed, the sharper the radiation cone, and the greater the deviation of the radiation direction from the particle speed direction.

The characteristics of ChR distinguish it from other radiation mechanisms like CR and ICS. The main differences are the direction of radiation and the effect of particle mass on radiation. Firstly, the CR direction is basically along the particle motion direction. The greater the velocity,  the radiation direction is closer to the velocity direction. In contrast, the radiation direction in the case of ChR could exhibit greater deviations compared to CR. Secondly, the mass of moving particles has minimal impact on radiation for ChR while the radiation power $\propto 1/m^2$ for the CR.  Besides, the formation mechanism of the bunch is simply discussed as follows. The outflow from the polar cap of the neutron star is probably unsteady. Due to the interaction between the near-surface parallel electric field and charged particles emitted from the surface, the inner gap of the neutron star could produce inhomogeneous sparks \citep{Ruderman1975,Zhang1996,Gil2000}. These sparks of electron-positron pairs would form bunches in the magnetosphere of a neutron star by two-stream instability \citep{Usov1987,Asseo1998,Melikidze2000}.  \cite{LuKumar2018}  proposed the bunches could form via two-stream instability in the twisted magnetosphere of magnetars before the magnetic reconnection is triggered. The plasma in the twisted magnetosphere of a magnetar is turbulent and clumpy owing to the two-stream instability.  When magnetic reconnection occurs, the pre-existing density clumps would provide charge bunches for the operation of the coherent mechanism \citep{LuKumar2018}. The electric currents flow along the magnetic field lines of a magnetar whenever the magnetic field lines are twisted by crustal motions. Besides, the two-stream instability causes density fluctuations on length scales longer than the effective plasma skin depth.  For the ChR,  the essential condition for the occurrence of two-stream instability is the velocity difference between the relativistic particle and the surrounding plasma. Considering a similar scenario, the large density fluctuations can develop on a length scale longer than the effective skin depth $\ell_{\text {skin }} \simeq 1.7 \times 10^{-2} \mathrm{~cm} \gamma_{c, 3}^{1 / 2} n_{18}^{-1 / 2}$, where number density $n_{18}$ and $\gamma_{c,3}$ are the fiducial parameters used in \cite{LuKumar2018}. Thus, the density fluctuations in these clumps could provide charge bunches for the ChR mechanism to operate. We summarize the properties of coherent CR, ICS, and ChR in Table \ref{table}.

Charged bunches continuously form and diffuse within the magnetosphere of a neutron star. Given the requirements for coherence and the need for the radiative process timescale to be shorter than the duration of the bunch existence, it is unlikely that the persistent bunches generated by the linear Langmuir wave can produce CR. However, some solutions could alleviate the conditions. For example, the observed frequency is significantly lower than the typical frequency associated with the emission process, the relationship between bunch length and the half-wavelength of the Langmuir wave can be disrupted by non-linear effects, and the charged bunches might exist for a duration much longer than the plasma oscillation period \citep{Melikidze2000}. Thus, a non-linear Langmuir wave is the preferred mechanism for generating the charged bunches that contribute to the coherent CR. Additionally, a fluctuating bunch could also make a contribution to the coherent radio emission. The emission spectrum by a single fluctuating bunch is significantly suppressed compared with that of a single persistent bunch, and the radiation by multiple fluctuating bunches is the incoherent summation of the emission by single bunches if the bunch separation is longer than the wavelength \citep{Yang2023}.

 \begin{figure*}
		\centering
		\includegraphics[width=1\textwidth]{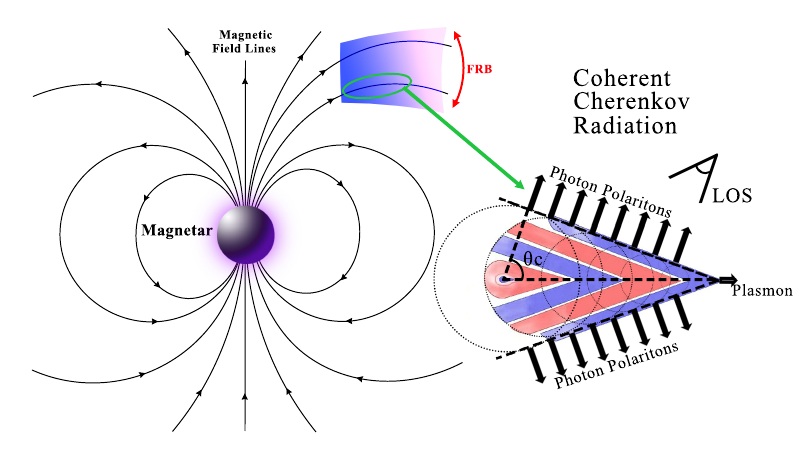}
		\caption{The geometric sketch of the FRBs produced by coherent ChR. Some relativistic particles are emitted from the polar cap of a magnetar and move along the field lines through a charge-separated magnetic plasma, emitting coherent ChR along the way. The relativistic particles lie on the top of the cone, and the black arrows denote the radiation direction. The electric vector lies in the plane formed by the particle velocity and radiation direction, perpendicular to the radiation direction.}
		\label{Fig1}
	\end{figure*}

\subsection{Magnetized Plasma Properties}
\label{sec2.1}

To determine whether ChR can be generated and to calculate the spectral distribution of radiation, it is necessary to know the refractive index of the medium first.  The refractive index formula can be derived from the dielectric tensor through the dispersion equation. The dispersion relation of electromagnetic waves propagating in a non-magnetized plasma can be derived by introducing a space and time variation of all quantities of the form exp $i(\mathbf{k} \cdot \boldsymbol{r} -\omega t)$ in Maxwell's equations and Newton's second law equation involving the Lorentz force, which can be given by $c k= n_r \omega$, where $\omega$ is the angular frequency,  $\mathbf{k}$ is the wave vector.  If a plasma carries an ordered magnetic field, it will significantly affect the propagation characteristics of electromagnetic waves and make the ChR possible. The existence of a magnetic field will cause spatial anisotropy in magnetized plasma. We consider a completely charge-separated plasma in a magnetic field $B$. The plasma particles move parallel to the magnetic field direction with velocity $\beta_p$ and energy $\gamma_p$ (in units of the vacuum velocity of light and the rest of energy, respectively). The dispersion equation for an electromagnetic wave of frequency $\omega^{\prime}$ in the plasma comoving frame, propagating through the plasma at an angle $\theta^{\prime}$ with respect to the field direction, is given by \citep{Heintzmann1975}
\begin{equation}
\begin{aligned}
&\frac{\omega^{\prime 2} \omega^2_B}{\omega^4_p}(n^{\prime 2}_r -1)\left[n^{\prime 2}_r \cos^{2}\theta^{\prime} -1-(n^{\prime 2}_r -1) \frac{\omega^{\prime 2}}{\omega^2_p}\gamma^2_p\Gamma\right]\\&=\Gamma (1- \frac{\omega^{\prime 2}}{\omega^2_p}\gamma^2_p\Gamma)\left[1+(n^{\prime 2}_r -1 )\frac{\omega^{\prime 2}}{\omega^2_p}\right]^2,
\end{aligned}
\end{equation}
where $\Gamma=(1-n^{\prime}_r\beta_p\cos\theta^{\prime})^2$, the relativistic plasma frequency $\omega_{p}=(4 \pi n_{e}^{\prime} e^{2}/\gamma_p m_{e})^{1/2}$, the relativistic electron gyration frequency $\omega_{B}=e B/\gamma_p m_{e} c$, $m_e$ is the mass of the electron, $n_r$ is the index of refraction. For a magnetar with surface magnetic field $B_{\rm s}=10^{15} {\rm G} B_{\rm s,15}$, and rotation period $P$, the Goldreich–Julian charge number density \citep{Goldreich1969,Ruderman1975} is $n_{\rm GJ}\simeq 6.9\times 10^7 {\rm cm}^{-3} B_{\rm s,15}P_0^{-1}R^{-3}_8$. We assume the stellar radius $R_{\rm s}=10^{6} {\rm cm}$, which is normalized. The plasma number density $n_e^{\prime}$ in the comoving frame can be estimated as $n_e^{\prime}=\xi n_{\rm GJ}$, where $\xi$ is the multiplicity parameter due to electron–positron pair production.
In the magnetosphere of a neutron star, the plasma is believed to satisfy the relation $\omega_p^2=2\xi\Omega \omega_B $, where $\Omega$ is the rotation frequency of the neutron star. The dispersion equation can be given by $n^{\prime 2}_r \cos^{2}\theta^{\prime} -1-(n^{\prime 2}_r -1) \frac{\omega^{\prime 2}}{\omega^2_p}\gamma^2_p (1-n^{\prime}_r\beta_p \cos\theta^{\prime})^2=0$.  We assume that relativistic charged particles with velocity $\beta$ and energy $\gamma_e$ is moving through the magnetized plasma in the same direction as the plasma particles. This particle will emit ChR when the condition of $n^{\prime}_r \beta \cos\theta^{\prime}=1$ is satisfied. We will give the index of refraction $n^{\prime}_r$ and the emission angle $\theta^{\prime}_c$ as functions of $\omega^{\prime}$  by the above condition and the solution of the dispersion equation, one has  \citep{Kolbenstvedt1977} 
\begin{equation}
n^{\prime 2}_r = 1+\left(\frac{\omega^{\prime}_c}{\omega^{\prime}}\right)^2,\label{nr}
\end{equation}and 
\begin{equation}
\sin^{-2}\theta^{\prime}_c = 1+\left(\frac{\omega^{\prime}}{\omega^{\prime}_c}\right)^2,\label{theta}
\end{equation}
where the characteristic frequency in the lab frame can be written as  
\begin{equation}
\begin{aligned}
\nu &=\frac{\omega_c}{2\pi}=\frac{1}{2\pi} \gamma_p\omega_p  \left|\gamma_e \gamma_p (\beta-\beta_p) \right |^{-1}\\ &\simeq 1 {\rm GHz} \;\xi^{1/2}_2 B^{1/2}_{\rm s,15}P_0^{-1/2}R^{-3/2}_8 \gamma_{p,1.5}^{-1/2} \gamma^{-1}_{e,2.5}, 
   \end{aligned}\label{frequency}
\end{equation}
$\omega_c=\gamma_p\omega^{\prime}_c$, and the index of refraction in the lab frame can be given by \citep{Melrose2004} 
\begin{equation}
n^{2}_{r}=1+\frac{n_r^{\prime 2}-1}{\gamma_p^2(1+n^{\prime}_r\beta \cos\theta^{\prime})^2}.
\end{equation}
As shown in Equation (\ref{nr}), when relativistic charged particles move through a magnetic plasma, the condition of $n^2_r > 1$ can be satisfied for generating ChR. However, when  $\omega\gg\omega_c$, it makes $n^2_r \simeq 1$, indicating that ChR cannot occur under this condition (also discussed in Section \ref{sec2.4}). The emitted frequency forms a hollow cone with an opening angle around the trajectory of the emitting particle. Notably, the cone's width decreases as the frequency increases, as described in Equation (\ref{theta}).
As demonstrated in Equation (\ref{frequency}), $\gamma_e$ and $\gamma_p$ remain relatively constant throughout the emission phase. Consequently, the characteristic frequency exhibits a limited range of values, suggesting its intrinsic narrowband mechanism. Furthermore, the frequency decreases with radius, giving a radius-to-frequency mapping for the ChR, which offers an explanation for the observed subpulse frequency down-drifting profile. We find that the characteristic frequency for CR $\sim 2.3 {\rm GHz}\; \gamma^3_{e,2.5}\rho^{-1}_8 $, where $\rho$ is the curvature radius. For the ICS, one has  $1{\rm GHz}\; \gamma^2_{e,2.5}\nu_{0,4}(1-\beta \cos\theta_i) $, where $\nu_0$ is the frequency of low-frequency electromagnetic waves, and $\theta_i$ is the angle between the incident photon momentum and the electron momentum. It is crucial to note that the ChR mechanism fundamentally relies on the dispersive properties of the plasma, distinguishing it from both the CR and ICS mechanisms. The CR mechanism illustrates the curved trajectories of charged particles during acceleration, while the ICS mechanism involves the acceleration of bunched particles via an oscillating electromagnetic field generated by low-frequency waves.  For the ChR mechanism, the electrons in the medium are accelerated when the field of the ``superluminal'' particle sweeps through the medium. In the following subsection, we proceed to calculate the emission power of ChR.

\subsection{ChR Power}
\label{sec2.2}

\begin{figure*}
		\centering
		\includegraphics[width=0.7\textwidth]{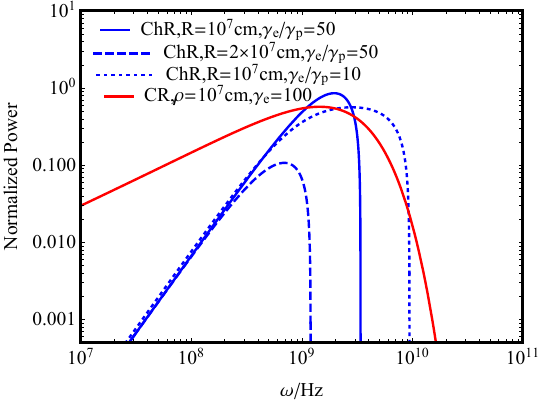}
	\caption{The normalized radiated power spectrum of an individual electron for CR and ChR. The solid red line denotes the CR ($\rho=10^7$cm, $\gamma_e=100$), and the blue lines denote the ChR. The solid blue line, blue dashed line, and blue dotted line indicate radiation radius $R= 10^7$ cm, $\gamma_e/\gamma_p=50$; $R=2\times10^7$ cm, $\gamma_e/\gamma_p=50$; and $R= 10^7$ cm, $\gamma_e/\gamma_p=10$; respectively. We adopt below physical parameters:  $\xi=100$, $B_s =10^{15}$G, $P=1$s. }
		\label{Fig2}
	\end{figure*}

In the following, we show the emission power of the ChR process. 
The emission power of a bunch in the lab frame: the emission power of an individual electron for ChR can be given by

\begin{equation}
    P_{e}^{\rm ChR}=\frac{\beta e^{2} }{c}\int_{\min}^{\max}\omega\left(1-\frac{1}{n^{2}_{r} \beta^{2}}\right)d\omega\simeq 1.0\times10^{-12}\;\rm{erg/s},
\end{equation}
where $\omega=\gamma_p \omega^{\prime}$. We adopt below physical parameters: $\omega_{\min}=10^7$ Hz, $\omega_{\max}=10^{10}$ Hz, $R=10^8$ cm, $\gamma_e=10^{2.5}$, $\gamma_p=10^{1.5}$, $\xi=100$, $B_{\rm s} =10^{15}$G, $P=1$s. Given that the Lorentz factor of the relativistic particle from the polar cap is generally greater than the  Lorentz factor of the plasma, we consider the above parameters.  The $\theta$ can be an arbitrary angle in the lab frame, so all the factors involving $\theta$ can be of the order of unity. ChR depends on the properties of the surrounding plasma environment. It is evident that the condition for ChR generation is met when $n^2_r > 1/\beta^2$. We find that the emission power of a single particle for CR $ P_{e}^{\rm CR}$ = $4.6\times 10^{-15} \; \gamma^4_{e,2.5}\rho^{-2}_8$. For the ICS, one has $P_{e}^{\rm ICS}$ $1.6\times 10^{-7} (\delta B_{0,6})^2 R^{-2}_8$, where $\delta B_{0}$ is the oscillation amplitude of the magnetic field of the electromagnetic waves and $R$ is the radius of the emission region.
The emission power of an individual electron $P_{e}^{\rm CR}<  P_{e}^{\rm ChR}<P_{e}^{\rm ICS}$, meaning that the required coherence for ChR to explain the high brightness temperature of FRB is significantly reduced compared to the CR. In Figure \ref{Fig2}, we present the normalized radiated power spectrum for an individual electron in CR and ChR. The plot suggests that the bunched coherent ChR mechanism has the advantage of generating a narrower-band spectrum compared to CR. While \cite{Yang2018} previously calculated a relatively broad spectrum for electron–positron pair bunches in coherent CR, it's important to note that charge separation can lead to a narrower-band spectrum, as shown in \cite{Yang2020}. Figure \ref{Fig2} illustrates the effect of the ratio of $\gamma_e/\gamma_p$ on the spectrum, represented by the solid and dotted blue lines. As this ratio increases, the spectrum becomes narrower. This suggests that when the velocity of plasma particles is much lower than that of beam particles, it is favorable for generating a narrower bandwidth.  Additionally, it's worth noting that the peak value of the energy spectrum is affected by the ChR radius, as evident in the solid and dashed blue lines shown in Figure \ref{Fig2}. It indicates that the superposition of radiation at different radius may contribute to a relatively broad energy spectrum. When the wave frequency is small compared with both $\omega_p$ and $\omega_B$, and the magnetic energy exceeds the plasma energy, the FMS and Alfv\'en waves would exist in this system. The FMS waves are a low-frequency $\omega\ll\omega_p$ limit of the X-mode of electromagnetic wave. As they propagate towards larger radius, the FMS waves are converted into the X-mode, which could eventually escape as outgoing electromagnetic waves \citep{Lyubarsky2021}.
We discuss the non-linear mode coupling will not prevent the escape of waves as follows. The interaction of FMS waves inevitably leads to the generation of Alfv\'en waves. These Alfv\'en waves evolve into a turbulent cascade, resulting in the dissipation of wave energy before they escape the magnetosphere. Hence, a weaker mode coupling strength could alleviate the non-linear interaction effect. The rate of the non-linear interaction could be estimated as \citep{Lyubarsky2021}
\begin{equation}
G \sim\left(\frac{\delta B}{B}\right)^2 \nu,
\end{equation}
where $\delta B/B=(L_{\rm FRB}/c)^{1/2} R^2/ \mu$, $ L_{\rm FRB}$ is the luminosity of FRB, $\delta B$ is the amplitude of the waves, $B$ is the background field, and $\mu$ is the magnetic momentum of the magnetar. If the characteristic time of the non-linear interaction $1/G$ is shorter than the characteristic propagation time, then most of the initial wave energy ultimately transforms into heat. In other words, the source is transparent to the non-linear interaction if the optical depth $\tau<1$. The
optical depth to the non-linear interaction can be given by  $\tau \simeq G R/c= 1.1\; L_{\rm FRB, 38} R_8^5 \mu_{33}^{-2} \nu_9$. One can see that the escape of waves is not significantly affected by non-linear interaction within the emission radius due to the small value of $\delta B/B$. Thus, only bursts with relatively low luminosity ($L < 10^{38}$ erg/s) can escape the magnetosphere within the framework of ChR.

ChR is a collective effect involving a multitude of particles within the medium.  The medium responds coherently to the passage of an ultrarelativistic primary beam. $N_e$ charged particles are divided into $N_{-}$ electrons and $N_{+}$ positrons. A bunch of $N_e$ electrons in the medium would emit a power $\sim (\eta N_e)^2 P_{e}^{\rm ChR}$, where $\eta$ denotes the fractional excess of negative
charges (see section \ref{sec2.4}). $N_e$ can be estimated by

\begin{equation}
N_{e}=V_{\rm b} n_{e} \simeq 1.9 \times 10^{22} \gamma_{e,2.5}^{3}\nu_{9}^{-3} \xi_2 B_{\rm s,15}P_0^{-1}R^{-3}_8,
\end{equation}
where $n_{e}=\gamma_{e} n_{e}^{\prime}$ in the lab frame, the coherent patch contributes to the FRB radiation for a time duration of the order of $\nu^{-1}=\lambda/c$ in the observer’s frame, $V_{\rm b}=A_{\rm b} \lambda$ is the maximum coherent volume in the lab frame, $A_{\mathrm{b} } \sim \pi\left(\gamma_{e} \lambda\right)^{2}$ is the cross-section of the bunch \citep{Kumar2017}. Considering the possible existence of $N_b$ independent bunches that contribute to the observed luminosity at an epoch, we write the total true luminosity (not isotropic equivalent) in the lab frame as
\begin{equation}
\begin{aligned} L & \simeq N_{b} (\eta N_{e})^{2}  P_{e}^{\rm ChR} \gamma_{e}^{2} \\ & \simeq 3.5 \times 10^{39} \mathrm{erg} \;\mathrm{s}^{-1} N_{b, 4} P_{e,-12}^{\rm ChR}\eta^{2}_{-1} \gamma_{e,2.5}^{8}\nu_{9}^{-6} \xi_2^2 B^2_{\rm s,15}P_0^{-2}R^{-6}_8, \end{aligned}\label{luminosity}
\end{equation}
which is consistent with the true FRB luminosity from the observations. The true luminosity of the model can be compared with the observed isotropic luminosity by considering
the beaming correction factor \citep{zhangbook2022}, so one has $L \simeq L_{\rm {iso}} \max (\pi \theta_j^2 ,\pi/ \gamma_e^2)$, where $\theta_j$ is the half opening angle of
the jet. When $\theta_j \leq \gamma_e^{-1} $, one can get  the isotropic luminosity $L_{\rm {iso}} \simeq 1.1 \times 10^{44} \mathrm{erg} \;\mathrm{s}^{-1}\;N_{b, 4} P_{e,-12}^{\rm ChR}\eta^{2}_{-1} \gamma_{e,2.5}^{10}\nu_{9}^{-6} \xi_2^2 B^2_{\rm s,15}P_0^{-2}R^{-6}_8$. The condition that the bunches constituting the pulse propagate along the magnetic field lines implies that the energy density of the local magnetic field must exceed the energy density of the plasma \citep{LuKumar2018}, i.e. $B \gtrsim 4.1 \times 10^{9} \mathrm{G}\; B_{\rm s, 15}^{-1 / 2} \zeta_1^{3 / 4}(L_{\text {iso,44}}/f_{\rm r})^{3 / 4}$, where $f_{\rm r}$ is the radiation efficiency in the radio band and the magnetic energy density exceeds that of the plasma by at least a factor of $\zeta \gg 1$.

We find that the coherent ChR can easily account for the typical FRB luminosity with a moderate number of $N_b\sim 10^4$ for the same electron Lorentz factor and emission radius (similar to the case for the coherent ICS). This is primarily due to the emission power of an individual electron $P_{e}^{\rm CR}<  P_{e}^{\rm ChR}<P_{e}^{\rm ICS}$.  As shown in Equation (\ref{luminosity}), the luminosity would be affected by the refractive index in the medium, which is significantly different from the coherent CR and ICS mechanisms.

\subsection{Required Electric Field }
\label{sec2.3}
To sustain FRB emission, a continuous electric field $E_\parallel$ is needed in the magnetosphere to continuously provide energy to the bunch. The scenario invoking CR by bunches was first proposed by \cite{Kumar2017}. Consider a bunch of $N_e$ electrons in the medium radiating coherently. The total energy can be given by $N_e \gamma_e m_e c^2 $, and the total emission power is $\eta^2 N^2_e  P_{e}^{\rm ChR}$. Therefore, the cooling timescale of the coherent ChR bunches can be calculated by 
\begin{equation}
\begin{aligned}
t_c&= \frac{N_e\gamma_e m_e c^2}{\eta^2 N^2_e  P_{e}^{\rm ChR}}\\&\simeq 1.4\times 10^{-12} {\rm s}\;\gamma_{e,2.5}^{-2}\nu_{9}^{3} \xi_2^{-1} B^{-1}_{\rm s,15}P_0 R^{3}_8 P_{e,-12}^{\rm ChR,-1}\eta^{-2}_{-1},
\end{aligned}
\end{equation}
which is much shorter than the typical FRB duration. Similar to the situation in the coherent CR mechanism, a parallel electric field is also needed to provide energy to the bunch continuously. The  parallel electric field can be calculated by $N_e e E_\parallel c = \eta^2 N^2_e  P_{e}^{\rm ChR}$, so one has 
\begin{equation}
\begin{aligned}
E_{\parallel}^{\rm ChR}&= \frac{\eta^2 N_e  P_{e}^{\rm ChR}}{ec}\\& \simeq 1.3\times 10^7 {\rm esu}\; \gamma_{e,2.5}^{3}\nu_{9}^{-3} \xi_2 B_{\rm s,15}P_0^{-1}R^{-3}_8 P_{e,-12}^{\rm ChR}\eta^2_{-1}.
\end{aligned}\label{Epar}
\end{equation}

The parallel electric field can sustain electrons moving with high Lorentz factors. Primary electrons can be accelerated from the polar cap surface to a high altitude in the slot gap, as this region can not be filled with electron–positron pairs generated from a pair production cascade  \citep{Muslimov2004}.
One may compare the parallel electric field required by the other two radiation mechanisms, the parallel electric field $E_{\parallel}^{\rm CR} \simeq 6.0\times10^6 {\rm esu}\; \gamma^7_{e,2.5}\rho^{-5}_8 \nu_{9}^{-3} \xi_2 B_{\rm s,15}P_0^{-1}$ for the coherent CR and  $E_{\parallel}^{\rm ICS} \simeq  8.6\times10^{10} {\rm esu}\; \xi_2 \nu^{-3}_9 \gamma^2_{e,2.5} B_{\rm s,15}P_0^{-1}f(\theta_i)\delta B^2_{0,6}R^{-5}_8 $ for the coherent ChR.  As shown in  Equation (\ref{Epar}), for the same magnetospheric condition, Lorentz factor, and emission radius, the magnitude of the parallel electric field is primarily  determined by the emission power of an individual electron.  Given that the emission power of an individual electron $P_{e}^{\rm CR}<  P_{e}^{\rm ChR}<P_{e}^{\rm ICS}$, it follows that $E_{\parallel}^{\rm CR}<  E_{\parallel}^{\rm ChR}<E_{\parallel}^{\rm ICS}$. Therefore, the parallel electric field $E_{\parallel}^{\rm ChR}$ can continuously supply energy to the bunch to power FRB emission.

\subsection{Coherence Factor }
\label{sec2.4}
The high brightness temperature of FRBs implies the necessity of coherent radiation. In this section, our primary focus is on calculating the coherence factor of ChR. As the medium responds coherently to the passage of a charged particle, these particles also radiate in a dielectric medium.  For a linear, homogeneous, and isotropic medium, the displacement field $\mathbf{D}=\epsilon\mathbf{E}$, the magnetic field strength $\mathbf{H}=(\mu)^{-1}\mathbf{B}$, where $\epsilon=\epsilon_r \epsilon_0$ and  $\mu=\mu_r \mu_0$ are the total permittivity and permeability. $\epsilon_r$ and $\epsilon_0$ are the relative and free permittivity, respectively. $\mu_r$ and $\mu_0$ are the relative and free permeability, respectively. The electric field of the Cherenkov radio pulse generated by a single charged particle moving in a dielectric medium can be expressed as \citep{Zas1992}

\begin{equation}
\mathbf{E}(\omega, \mathbf{x})=\frac{e \mu_r i \omega}{2 \pi \epsilon_0 c^2} \frac{e^{i k R}}{R} e^{i(\omega-\mathbf{k} \cdot \mathbf{v}) t_1} \mathbf{v}_{\perp}\left[\frac{e^{i(\omega-\mathbf{k} \cdot \mathbf{v}) \delta t}-1}{i(\omega-\mathbf{k} \cdot \mathbf{v})}\right],
\end{equation}
where $e$ is the charge of an electron, $\mathbf{v}$ is the particle velocity, $\mathbf{v}_{\perp}$ is the projection of the velocity onto a plane perpendicular to the direction of observation, and $\delta t=t_2 -t_1$ is the time interval between the endpoints of the track. $R$ represents the retarded distance, measured from the point on the particle's trajectory where the radiation was emitted to the observation point. The charged particle moves uniformly between two fixed points denoted by subscripts 1 and 2. When observation angles are in close proximity to the Cherenkov angle, the electric field can be expressed as

\begin{equation}
\mathbf{E}(\omega, \mathbf{x})=\frac{e \mu_r i \omega}{2 \pi \epsilon_0 c^2} \frac{e^{i k R}}{R} e^{i(\omega-\mathbf{k} \cdot \mathbf{v}) t_1} \mathbf{v}_{\perp}\delta t.
\end{equation}
To simplify the calculations, we consider the electric field amplitude unity, and the fields are polarized in the same direction. Additionally, we assume a dielectric medium composed of $N_e$ charged particles, which can be further divided into $N_{-}$ electrons and $N_{+}$ positrons. Given that the electric fields of electron and positron have an opposite sign for the coherent ChR, the total electric field amplitude $E$ at the detector can be written as  \citep{Conti2017}  
\begin{equation}
E=\sum_{i=1}^{N_{-}}e^{i(\Phi_i-\omega t)}-\sum_{i=1}^{N_{+}}e^{i(\varTheta_i-\omega t)},
\end{equation}where $\Phi_i$ and $\varTheta_i$ are the phases of the electromagnetic wave emitted by the $i$-th electron and positron, respectively. $\Phi_i$ and $\varTheta_i$ are independent random variables with the same distribution for any $i$ (i.e., $f( \Phi_i)$=$f(\varTheta_i)$, where $f$ is the probability density distribution function). The average values are the same for any $i$ (i.e., $\langle \Phi_i\rangle$=$\langle \varTheta_i\rangle$). We can get the energy $W$, one has
\begin{equation}
\begin{aligned}
W=E E^{*}&=\sum_{i, j=1}^{N_{-}} e^{i\left(\Phi_i-\Phi_j\right)}+\sum_{i, j=1}^{N_{+}} e^{i\left(\Theta_i-\Theta_j\right)}\\
&-2 \sum_{i=1}^{N_{-}} \sum_{j=1}^{N_{+}} \cos \left(\Phi_i-\Theta_j\right).
\end{aligned}
\end{equation}
The $N$ terms with $i=j$ correspond to incoherent emission from each charge. In this model,  coherent emission is described by the $N(N-1)$ terms with $i\neq j$. The average energy value can be simplified further, as demonstrated in \citep{Conti2017}
\begin{equation}
\langle W\rangle=\left(N_{-}+N_{+}\right)+\left[\left(N_{-}-N_{+}\right)^{2}-\left(N_{-}+N_{+}\right)\right]\langle\cos \Phi\rangle^{2}.
\end{equation}The factor $\langle\cos \Phi\rangle^{2}$ determines the coherence of particles. The phase $\Phi$ is uniformly distributed between $-\Phi_0$ and $\Phi_0$, one has
\begin{equation}
\langle\cos \Phi\rangle=\frac{1}{2 \Phi_{0}} \int_{-\Phi_{0}}^{\Phi_{0}} \cos \Phi d \Phi=\frac{\sin \Phi_{0}}{\Phi_{0}}.
\end{equation}
For $\Phi_0 \gg1 $, we obtain $\langle\cos \Phi\rangle=0$, and the total energy is $\langle W\rangle=N_{-}+N_{+}$= $N_e$, representing a scenario of total incoherence. For $\Phi_0 \rightarrow 0$, the total energy is given by $\langle W\rangle=\left(N_{-}-N_{+}\right)^2=(\eta N_e)^2$, indicating complete coherence. The above results are consistent with the calculations by \cite{Melrose2017} about the bunched coherent ChR (i.e., coherent emission from a bunch treated as a collection of discrete particles). It's important to note that the coherence factor of ChR significantly differs from that of CR and ICS. The emission power of the bunch for CR and ICS can be estimated by $N^2_e P_e$ \citep{Yang2018,ZB2022}.

\begin{figure}
		\centering
		\includegraphics[width=0.48\textwidth]{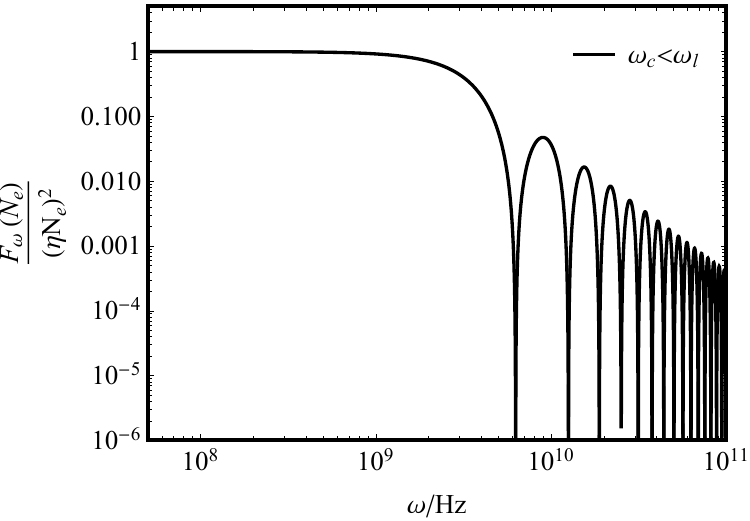}
        \includegraphics[width=0.48\textwidth]{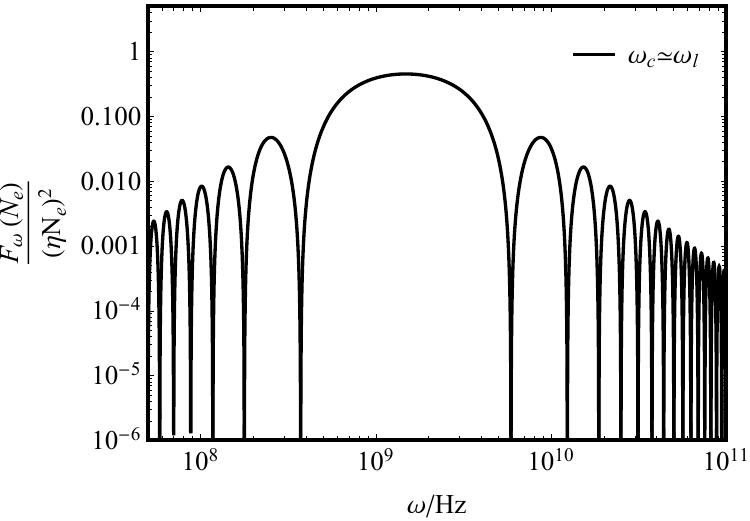}
        \includegraphics[width=0.48\textwidth]{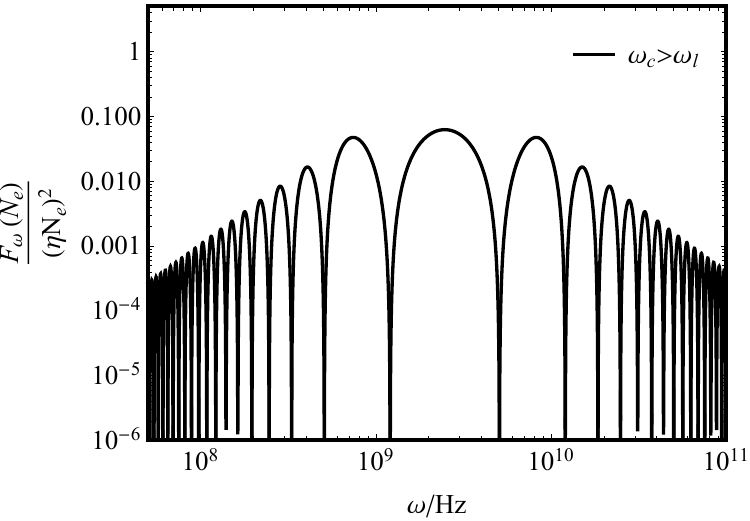}
		\caption{The coherence factor as a function of frequency for ChR. The upper, middle, and bottom panels denote $\omega_c<\omega_l$, $\omega_c\simeq\omega_l$, and $\omega_c>\omega_l$, respectively. }
		\label{Fig3}
	\end{figure}

Assuming that the trajectories of $N_e$ electrons are the same but the electrons are injected at different times. The retarded position of the $j$th electron can be given by $\boldsymbol{r}_j(t)=\boldsymbol{r}(t)+\Delta \boldsymbol{r}_j(t)$, where $\boldsymbol{r}(t)$ denotes the retarded position of the first electron, and $\Delta \boldsymbol{r}_j(t)$ denotes the relative displacement between the first electron and the $j$th electron. The total energy radiated per unit solid angle per unit frequency interval for the ChR can be approximately written by (see Appendix)
\begin{equation}
\begin{aligned}
\frac{dI}{d\omega d\Omega} &=\frac{e^2 \omega^2}{4\pi^2 c }n_r(\omega) \left| \int_{-\infty}^{\infty}\boldsymbol{n}\times (\boldsymbol{n}\times \boldsymbol{\beta}) e^{i\omega (t - n_r(\omega)\boldsymbol{n}\cdot \boldsymbol{r}(t) /c)} dt \right|^2  \\ &\times   \left|\sum^{\eta N_{e}}_j e^{-i\omega (n_r(\omega) \boldsymbol{n}\cdot \Delta \boldsymbol{r}_j /c)}\right|^2,
\end{aligned}
\end{equation}where the coherence factor
\begin{equation}
\begin{aligned}
F_\omega (N_e)&=\left|\sum^{\eta N_{e}}_j e^{-i\omega (n_r(\omega)\boldsymbol{n}\cdot \Delta \boldsymbol{r}_j /c)}\right|^2 \\
&= \left|\frac{\eta N_{e}}{L}\int_0^L e^{-i\omega x \cos\varphi n_r(\omega)/c} dx\right|^2\\
&=({\eta N_{e}})^2 \left| \frac{1- e^{-i\omega L \cos\varphi n_r(\omega)/c}}{i\omega L \cos\varphi n_r(\omega)/c}\right|^2,
\end{aligned}
\end{equation}
where $L$ is the bunch length, the coherence factor shows that the separation between bunches is much smaller than the wave length \citep{Yang2023}.  We define $\omega_l \simeq 2c/L$ and $\omega_c$ as the characteristic frequency of the bunch length \citep{Yang2018} and the characteristic frequency of ChR (as shown in Equation (\ref{frequency})), respectively.  As shown in the upper panel of Figure \ref{Fig3} ($\omega_c<\omega_l$), when the frequency of the electromagnetic waves is significantly lower than the characteristic frequency of a bunch, it suggests that the emission from each electron exhibits nearly the same phase, thus allowing for coherence in this scenario. Besides, coherence gradually decreases with frequency when $\omega>\omega_l$. As shown in the middle panel of Figure \ref{Fig3} ($\omega_c\simeq\omega_l$), coherence can occur at the frequency around $\omega_c$. However, if $\omega_c>\omega_l$, as shown in the bottom panel of Figure \ref{Fig3}, the emission becomes incoherent.

\section{Observation Test of the Emission Mechanisms}
\label{sec3}

\subsection{Spectrum}
\label{sec3.1}
The spectrum provides an effective way for testing the emission mechanism, and many observations have suggested that FRBs have a narrow intrinsic spectrum. The spectral shape generally can be assumed as a power law function $F_\nu\propto \nu^{-\alpha}$. For instance, the power law index $\alpha=4\pm1$ for the Lorimer burst \citep{Lorimer2007},  $\alpha=7.8\pm0.4$ for the FRB 110523 \citep{Masui2015}, and $\alpha=-10.4$ to $+13.6$ for the FRB 121102 \citep{Spitler2016}. Additionally, observations of over 700 bursts from FRB 20201124A detected by FAST also indicate that the majority of repeating FRBs exhibit a narrow spectrum \citep{Zhou2022}. We discuss the properties of the spectrum for the different emission mechanisms as follows. From Figure \ref{Fig2}, we show the spectral power per unit frequency interval for coherent ChR and the energy radiated per unit frequency interval per unit solid angle for CR. Equation (\ref{ap3}) in the Appendix reveals that ChR exhibits significant directionality due to the presence of a delta function. A characteristic frequency for ChR can be given by $\omega_c = \gamma_p \omega_p  \left|\gamma_e \gamma_p (\beta-\beta_p) \right |^{-1} $ in the rest frame of the particle bunch. The nearly constant values of these parameters during the emission phase indicate that the spectrum generated by the coherent ChR mechanism is indeed narrow. Besides, it is worth noting that the spectral power experiences a rapid increase with frequency. During the rising stage, the spectral index of the ChR spectrum is steeper than that of CR, as shown in Figure \ref{Fig2}. For coherent CR, the low-frequency index of the spectrum is typically 2/3 \citep{Yang2018}. Furthermore, We find that the power spectrum drops rapidly after reaching the peak, and notably, the spectral index of the ChR spectrum during the descending phase is steeper compared to that of CR. Thus, this spectrum profile holds potential for testing the emission mechanism.

In the case of coherent CR, the spectrum has been identified as a broken power law, distinguished by certain characteristic frequencies \citep{Yang2018}. A broad intrinsic spectrum ($\Delta \nu /\nu \sim 1$) can be emitted by coherent bunches, but the absorption of low-frequency radio emission \citep{Yang2018} would produce an observed narrower spectrum. Moreover, the narrow spectrum could also be generated by the charge separation within the framework of CR \citep{Yang2020}. On the other hand, for the coherent ICS, the characteristic frequency has a narrow value range (i.e., the oscillation frequency, the Lorentz factor, and the photon incident angle vary very little), indicating that it is inherently a narrow-band mechanism \citep{ZB2022}.

\subsection{Frequency Drifting }
\label{sec3.2}

The observed time-frequency behavior is characterized by a downward drifting trend in the majority of repeating FRBs \citep{Gajjar2018,Michilli2018,Hessels2019,Josephy2019,Caleb2020,Day2020,Platts2021}. 
This phenomenon appears to be a common feature among repeaters, with the best-fit model for all FRBs yielding a drifting power-law index of $2.45\pm0.21$, and for FRB 121102, a drifting power-law index of $2.29\pm0.36$ \citep{WangYang2021}. Some efforts have been made to explain the drifting pattern \citep{Cordes2017,Wang2019,Lyutikov2019,Margalit2020,Rajabi2020,Tuntsov2021,Mahlmann2022}. The ``downwards drifting'' structure can be derived from Equation \ref{frequency} within the framework of ChR. Specifically, by calculating $d\nu/dR$, we find that it is proportional to $-R^{-5/2}$, leading to a drifting relationship expressed as $d\nu/dt \propto -\nu^{5/3}$. The predicted drifting rate index for coherent ChR is $5/3$, indicating that a higher-frequency spark is observed at an earlier time, consistent with observations. It's worth noting that the time-frequency downward drifting is a natural consequence of coherent CR, as discussed in \cite{Wang2019}. The drifting profile of FRBs reflects the frequency variation with location within the magnetosphere. In essence, a spark observed at an earlier time with a higher frequency is emitted in a more-curved magnetic field line.  The drift rate as a function of frequency can be given by $\dot{\nu}\propto -\nu^{2}$ within the framework of CR \citep{WangYang2021}. Therefore, the time-frequency structure of FRBs becomes a valuable tool for testing and understanding the radiation mechanisms.

\subsection{Polarization}
\label{sec3.3}
Polarization is a crucial probe for studying emission mechanisms and propagation processes \citep{qu2023}. Most FRBs display near 100$\%$ LP fractions and a flat PA across each pulse \citep{DS2021,Nimmo2021,CHIME2019,Gajjar2018,Michilli2018}. However, there are exceptions, such as FRB 180301, which display partial linear or circular polarization, along with significant variations in PAs over time  \citep{Luo2020,Cho2020}. Notably, an important CP has recently been observed in some FRB repeaters, including FRB 20201124A, FRB 20121102A, and 20190520B \citep{FZ2022,jiang2022,Hilmarsson2021,Kumar2021,XuH2021}.  
The observed high CP fraction may be attributed to either intrinsic radiation mechanisms \citep{WangJiang2021,WangYang2021,Tong2022,liu2023} or propagation effects, such as cyclotron resonance absorption and Faraday conversion \citep{Beniamini2022,qu2023}. Within the framework of CR, the emission would be highly linearly polarized if the opening angle of the bunch $\varphi_t \gg 1/\gamma$. However, if the particles of bunches are non-uniformly distributed or $\varphi_t \sim 1/\gamma$, the circularly polarized bursts would be generated within the emission cone \citep{WangJiang2021}. In the case of coherent ICS, the outgoing photon is linearly polarized after the ICS by the individual particle \citep{ZB2022}. CP can be generated by considering a geometrically asymmetric bunch \citep{qu2023}. 

Furthermore, we will discuss the polarization features of coherent ChR. The ChR is emitted in a narrow cone known as the Cherenkov angle by ultra-relativistic bunched particles. This cone has an opening angle of $\sim 1/\gamma$ in the direction of the particle's velocity. The radiation from a single charge can be described by two orthogonal polarized components ($A_{\parallel}$ and $A_{\perp}$).  The linearly polarized component is expressed as $L =\sqrt{Q^2+U^2}$, where $Q$ and $U$ are the stokes parameters. As shown in Equation (\ref{ap4}) of the Appendix, the component $A_{\perp} \sim 0$, indicating that the ChR  from a single fast particle is strongly linearly polarized. The electric vector of this linearly polarized radiation lies in the plane formed by the particle's velocity and the direction of radiation, and it is perpendicular to the direction of radiation. Similar to the CR, the pulses show a $\sim 100\%$ LP degree when the line of sight (LOS) sweeps at the center of the bunches (this case is similar to the emission from a single charge). The highly linearly polarized bursts are generally consistent with most FRB observations \citep{CHIME2019,Gajjar2018,Michilli2018}. Additionally, the CP fraction and the variation of PA might be influenced by the geometry of bunches and the on/off beam scenario \citep{WangJL2022}. Many propagation effects within the magnetosphere might also contribute the high CP fraction, primarily including the resonant cyclotron absorption, wave mode coupling, and wave propagation through the quasi-tangential regions \citep{WLH2010}. We plan to conduct further investigations into the polarization characteristics of ChR in our future work.

\section{Conclusions and Discussion}
\label{sec4}

In this paper, we have investigated coherent ChR as the potential radiation mechanism responsible for FRBs. Our study considers the scenario in which relativistic particles are emitted from the polar cap of a magnetar and move along magnetic field lines through a charge-separated magnetic plasma, emitting coherent ChR along their path. We have reached the following key conclusions:

1. The condition of $n^2_r > 1$ can be satisfied for coherent ChR production in the magnetospheric environment, providing a feasible mechanism for FRBs.

2. We have calculated the emission power of an individual electron for ChR $P_{e}^{\rm ChR}\simeq 1.0\times 10^{12}$ erg/s, and estimated a total luminosity $\simeq 3.5\times10^{39}$ erg/s. Importantly, the bunched coherent ChR mechanism offers the advantage of producing a narrower-band spectrum compared to other mechanisms.

3. We have calculated the coherence factor of ChR, which can be estimated as $(\eta N_e)^2  P_{e}^{\rm ChR}$. This factor significantly differs from those associated with the CR and ICS mechanisms.  Our analysis indicates that ChR can exhibit coherence when $\omega_c\lesssim \omega_l$, implying that emissions from individual electrons have nearly the same phase. This reduced coherence requirement compared to CR is advantageous in explaining the high brightness temperature of FRBs.

4. Coherent ChR produces a frequency downward-drifting pattern and emits $\sim100\%$ linearly polarized emission, which is consistent with observations of repeating FRBs.

 Furthermore, we acknowledge that considering the non-linear interaction effect may affect the escape of bursts from the magnetosphere. In this regard, it is plausible  that only relatively faint bursts ($L < 10^{38}$ erg/s) could escape the magnetosphere within the framework of ChR. The influence of non-linear interaction effect on wave propagation needs to be further investigated in detail.

Interestingly, even for different bursts from the same repeater FRB 121102, $\alpha$ varies from $-10.4$ to $+13.6$  \citep{Spitler2016}. This variability can be explained within the framework of coherent ChR. As shown in Figure \ref{Fig2}, the spectral index of the ChR spectrum during the rising and descending stage is steeper than that of CR. We have found that an increase in the ratio of $\gamma_e/\gamma_p$ results in a narrower spectrum. Variations in this ratio among different bursts from the same repeater could account for differences in the spectral index.

 Very recently, \cite{Yang2023} discussed the spectral characteristics of coherent CR, particularly the effects of fluctuating bunches, revealing a quasi-white noise component spanning a broader frequency band. Similarly, the spectrum of ChR may exhibit white-noise features due to bunch fluctuations. Besides, it's worth noting that the emission spectrum generated by a single fluctuating bunch is suppressed when compared to a single persistent bunch in CR \citep{Yang2023}.  The characteristic frequency of the ChR could modulate the emission spectrum. We will present the spectral features of coherent ChR by the fluctuating bunches in our future work.

	\acknowledgments
	
We are grateful to Fa-Yin Wang, Bing Zhang and an anonymous referee for helpful discussions and constructive suggestions. This work was supported by the National SKA Program of China (grant No. 2020SKA0120300) and the National Natural Science
	Foundation of China (grant No. 11833003).
	W.-Y.W is supported by a Boya Fellowship and the fellowship of China Postdoctoral Science Foundation No. 2021M700247 and No. 2022T150018.
	Y.-P.Y is supported by National Natural Science Foundation of China (grant No. 12003028), and the National SKA Program of China (2022SKA0130100). 
 G.-J.J is supported by National Natural Science Foundation of China (Grant Nos. 12273113 and 12041306),  the CAS Project for Young Scientists in Basic Research (Grant No. YSBR-063), and the Youth Innovation Promotion Association (2023331).

\begin{table*}
\centering
\caption{The properties of coherent CR, ICS, and ChR, including the emission power of an individual electron, luminosity, emission radius, frequency, parallel electric field, spectrum, coherence factor, linear polarization, and frequency drifting.}
\begin{tabular}{c|c|p{5cm}<{\centering}|p{5cm}<{\centering}}
\hline
 &Coherent Curvature Radiation  & Coherent Inverse Compton Scattering  &Coherent Cherenkov Radiation \\
\hline
\hline

Emission Power (\rm{erg/s}) &  $4.6\times 10^{-15}  \gamma^4_{e,2.5}\rho^{-2}_8$ & $1.6\times 10^{-7} (\delta B_{0,6})^2 R^{-2}_8$ & $1.0\times10^{-12}(R=10^8{\rm cm})$\\
\hline
Luminosity(erg/s) &  \makecell*[l]{$1.6 \times 10^{39} N_{b,4} \gamma_{e,2.5}^{12}$\\$\nu_{9}^{-6} \xi_2^2 B^2_{\rm s,15}P_0^{-2}R^{-8}_8$}  &  \makecell*[l]{$ 7.3\times 10^{44} \xi_2^2 N_{b,4}\gamma^8_{e,2.5}\nu^{-6}_9$  \\$B^2_{\rm{s},15}P_0^{-2} f(\theta_i) \delta B^2_{0,6}R^{-8}_{8} $} & \makecell*[l]{$3.5 \times 10^{39} N_{b,4} P_{e,-12}^{\rm ChR} \eta^{2}_{-1}$\\$ \gamma_{e,2.5}^{8}\nu_{9}^{-6} \xi_2^2 B^2_{\rm s,15}P_0^{-2}R^{-6}_8$} 
\\
\hline 
Emission Radius(cm) & $10^7\sim10^8$ & $10^7\sim10^8$or $10^8\sim10^9$  & $10^7\sim10^8$ \\
\hline
Frequency(GHz) & $2.3 \gamma^3_{e,2.5}\rho^{-1}_8$ & 1$ \gamma^2_{e,2.5}\nu_{0,4}(1-\beta \cos\theta_i) $ &\makecell*[l]{ $1 \xi^{1/2}_2 B^{1/2}_{\rm s,15}P_0^{-1/2}$\\$R^{-3/2}_8 \gamma^{-1/2}_{p,1} \gamma^{-1}_{e,2.5}$} \\
\hline
Parallel Electric Field(esu)& \makecell*[l]{$6.0\times10^6 \gamma^7_{e,2.5} \rho^{-5}_8$\\$ \nu_{9}^{-3} \xi_2 B_{\rm s,15}P_0^{-1}$} &  \makecell*[l]{$8.6\times10^{10}  \xi_2 \nu^{-3}_9 \gamma^2_{e,2.5}$\\$ B_{\rm s,15}P_0^{-1}f(\theta_i)\delta B^2_{0,6}R^{-5}_8 $} &  \makecell*[l]{$1.3\times 10^7  \gamma_{e,2.5}^{3}\nu_{9}^{-3}$\\$ \xi_2 B_{\rm s,15}P_0^{-1}R^{-3}_8 P_{e,-12}^{\rm ChR}\eta^2_{-1}$}  \\
\hline
Spectrum & wide & narrow & narrow\\
\hline
Coherence Factor & $N^2_e$ & $N^2_e$ & $(\eta N_e)^2$ \\
\hline
Linear Polarization & $\sim 100\%$ & $\sim 100\%$ &$\sim 100\%$ \\
\hline
Frequency Drifting & downward drifting & downward drifting &downward drifting \\
\hline
\end{tabular}\label{table}
\end{table*}

\appendix

We consider an electron moving along a trajectory $\boldsymbol{r}(t) $. The observation point is assumed to be far enough away from the acceleration region. The energy radiated per unit solid angle per unit frequency interval is given by  \citep{jackson1998,Rybicki1979}

\begin{equation}
\begin{aligned}
\frac{dI}{d\omega d\Omega} &=  \frac{e^2 \omega^2}{4\pi^2 c }\left| \int_{-\infty}^{\infty}\boldsymbol{n}\times (\boldsymbol{n}\times \boldsymbol{\beta}) e^{i\omega (t - \boldsymbol{n}\cdot \boldsymbol{r}(t) /c)} dt \right|^2 , \label{ap1}
\end{aligned}
\end{equation}where $\boldsymbol{\beta}$ is the dimensionless velocity and $\boldsymbol{n}$ is the unit vector between the electron and the observation point. The above equation is only applicable to electron radiation in a vacuum. We consider the radiation of electrons in the plasma with refractive index $n_r(\omega)$, and the above radiation spectrum needs to be modified. We make the simplifying assumption that particle motion occurs at a constant velocity. The equation for particle motion at constant velocity can be expressed simply as $\rho(t)=vt$. Thus, the energy radiated per unit solid angle per unit frequency interval in the plasma with refractive index $n_r$ can be given by
\begin{equation}
\begin{aligned}
\frac{dI}{d\omega d\Omega} =  \frac{e^2 \omega^2}{4\pi^2 c }n_r \left| \int_{-\infty}^{\infty}\boldsymbol{n}\times (\boldsymbol{n}\times \boldsymbol{\beta}) e^{i\omega (t - n_r\boldsymbol{n}\cdot \boldsymbol{\rho}(t) /c)} dt \right|^2. \label{ap2}
\end{aligned}
\end{equation} 
We define $\theta$ as the angle between radiation direction and velocity direction and then use the expression $\delta(x)=\frac{1}{2\pi}\int_{-\infty}^{\infty} e^{i\alpha x}d\alpha$, the above equation can be simplified as
\begin{equation}
\begin{aligned}
\frac{dI}{d\omega d\Omega} = \frac{e^2 n_r\beta^2 \sin^2\theta}{c} \left| \delta(1-n_r \beta \cos\theta)\right|^2.\label{ap3}
\end{aligned}
\end{equation} It can be seen from the equation that the radiation has significant directivity. The radiation can be generated within the Cherenkov angle $\theta_c$, where $\cos\theta_c=1/n_r \beta$.
Since the radiation or collision loss, we assume that the effective motion time of particles is 2$T$, then the integral limits for $t$ are $-T$ and $T$, one has
\begin{equation}
\begin{aligned}
\frac{dI}{d\omega d\Omega} = \frac{e^2 n_r\beta^2 \sin^2\theta}{c} \left( \frac{\omega^2 T^2}{\pi^2} \frac{\sin^2[\omega T (1-n_r\beta \cos\theta)]}{[\omega T (1-n_r\beta \cos\theta)]^2} \right).\label{ap4}
\end{aligned}
\end{equation} If $\omega T \gg 1$, the integral value reaches a maximum at the Cherenkov angle $\theta_c$, but it is not infinite. By integrating all the solid angles in the above equation, we can obtain the spectral power of the particle at the unit frequency interval near $\omega$, which can be written as
\begin{equation}
    P_e(\omega)=\frac{\beta e^{2} \omega}{c}\left(1-\frac{1}{n_r^{2} \beta^{2}}\right),\label{ap5}
\end{equation}it can be seen that the condition for generating Cherenkov radiation is when $n^2_r > 1/\beta^2$.
The total radiated power $P_e$ is obtained by integrating the frequency of the above equation. 

\end{document}